# Frequency-Aware Masked Autoencoders for Human Activity Recognition using Accelerometers

Niels R. Lorenzen[1,2,3], Poul J. Jennum[2], Emmanuel Mignot[3], Andreas Brink-Kjaer[1]

*Abstract*— **Wearable accelerometers are widely used for continuous monitoring of physical activity. Supervised machine learning and deep learning algorithms have long been used to extract meaningful activity information from raw accelerometry data, but progress has been hampered by the limited amount of labeled data that is publicly available. Exploiting large unlabeled datasets using self-supervised pretraining is a relatively new and underexplored approach in the field of human activity recognition (HAR).**

**We used a time-series transformer masked autoencoder (MAE) approach to self-supervised pretraining and propose two novel spectrogram-based loss functions: the log-scale mean-magnitude (LMM) and log-scale magnitude variance (LMV) losses. We compared these losses with the mean squared error (MSE) loss for MAE training. We leveraged the large unlabeled UK Biobank accelerometry dataset (*n* = 109k) for pretraining and evaluated downstream HAR performance using a linear classifier in a smaller labelled dataset.**

**We found that pretraining with the LMM loss improved performance compared to an MAE pretrained with the MSE loss, with 12.7% increase in subject-wise F1 score when using linear probing. Compared with a state-of-the-art ResNet-based HAR model, our LMM-pretrained transformer models performed better (+9.8% F1) with linear probing and comparably when fine-tuned using an LSTM classifier. The addition of the LMV to the LMM loss decreased performance compared to the LMM loss alone. These findings establish the LMM loss as a robust and effective method for pretraining MAE models on accelerometer data for HAR and show the potential of pretraining sequence-based models for free-living HAR.**

*Clinical Relevance*— **Improved algorithms for human activity recognition, enables more accurate monitoring of physical activity, which is crucial for assessing mobility, rehabilitation progress, and chronic disease management.**

## I. Introduction

Wearable accelerometry is a cost-effective tool for out-of-clinic health monitoring, with applications in human activity recognition (HAR) [1], [2], [3], [4], gait estimation [5], [6], and sleep monitoring [7]. The usefulness of these devices in understanding associations between different activities and health relies heavily upon the quality of algorithms to detect such activities based on the data.

Until recently, supervised classical machine learning and deep learning have been state-of-the-art for HAR using wearable accelerometers [8], [9], [10], [11]. However, since the arrival of foundation models for images and time-series relying on self-supervised pretraining [12], [13], multiple studies have employed different methods of self-supervised learning to wearable accelerometry data [1], [2], [4], [5], [14], [15]. However, only a few have done so with a large-scale dataset using the UK Biobank cohort [1], [5]. These both used multi-task learning based on a head-to-head comparison of masked reconstruction, multitask, and contrastive learning with a small amount of pretraining data [3]. However, the implementation of the masked autoencoder (MAE) in [2], [3] leaned heavily on the original implementation of the BERT language model [16], missing possible improvements from more recent advances in MAEs for time series such as higher masking rates, patching, and rotational positional embeddings [13], [17], [18]. Furthermore, using frequency-domain information has proven useful for supervised HAR models [8], [9], [10]. Other work has shown that integrating frequency awareness into masked-autoencoders for other bio-signals led to better and more generalizable representations [19].

In this work, we introduce frequency awareness to a time-domain MAE by adapting a spectrogram loss developed for audio signals [20]. We explore how variations to this spectrogram loss affect convergence and fine-tuning performance in HAR. We show that a modern transformer MAE trained with this loss can perform on par with or better than the current state-of-the-art in HAR when pretrained using the large UK biobank cohort.

## II. Methods

### A. Datasets

The unlabeled high-resolution wrist-worn accelerometer dataset from the UK Biobank [21] was used for self-supervised pretraining in this work. The dataset consists of 115,390 recordings of ~7 days of free-living wrist-worn accelerometry from 103,618 participants. The data was recorded with 100 Hz sampling rate on Axivity AX3 devices. The sequence-aware nature of our transformer-based model calls for evaluation using labeled continuous recordings of free-living data rather than short scripted sequences of single behaviors as we expect the model to learn natural transitions between behaviors. This severely limits the amount of publicly available datasets for downstream performance evaluation. For this purpose, we used the Capture-24 cohort [22] consisting of 151 participants with ~24 hours of data sampled at 100 Hz from Axivity AX3 devices. The data in this cohort has annotated day-time activities based on body-cam recordings and sleep periods based on a sleep-diary. The labels are reduced to the six categories bicycling, walking, mixed, vehicle, sit-stand, and sleep following previous research [1], [10].

[1] Department of Health Technology, Technical University of Denmark, Lyngby, Denmark.

[2] Danish Centre for Sleep Medicine, Copenhagen University Hospital-Rigshospitalet, Glostrup, Denmark.

[3] Department of Psychiatry and Behavioral Sciences, Stanford University, CA, USA.

## B. Preprocessing

Preprocessing included applying a 15Hz low-pass filter and resampling to 30Hz. The accelerometry data was then calibrated using the algorithm described in [23] without temperature correction. Stationary periods were detected using a 10 second moving standard deviation threshold of 0.013 g. Stationary periods of more than 90 minutes were classified as non-wear and dropped. Contiguous wear segments of at least 24 hours for the UK Biobank data and 1 hour for Capture-24 were kept. We excluded recordings without at least one period of the minimum wear time and those where calibration failed. This left 108,933 recordings from the UK biobank, and 149 from Capture-24.

## C. Model Architecture

In the model (Fig. 1a) each 50-minute input signal is first segmented along the time dimension into 300 non-overlapping 10-second patches. The patches are then flattened along the channel dimension and projected, in parallel, by a linear layer into 256-dimensional embedding vector. The 300 embeddings are then passed through 12 transformer encoder blocks. Each encoder block applies rotational embeddings (RoFormer) [18] and multi-head self-attention [24] with 8 attention heads, before a feed-forward layer with SwiGLU activations [25] and 682 hidden units. A residual connection with root-mean-square normalization [26] is applied *before* [27] each attention and feed-forward layer. The models have a total of ~10M parameters. During pretraining, a random subset of the embeddings was replaced with a mask token, which is a learnable parameter vector, and a linear layer acts as a reconstruction head, projecting the output feature vectors back into signal space. During fine-tuning, no masking was applied to the vector sequence, and the reconstruction head was replaced with a linear classification head.

## D. Pretraining

A 90:10 train-validation split was used for pretraining. Data was loaded in 24hrs of contiguous wear accelerometry data from 8 recordings. These were split into 32 50-minute windows with 5 minutes of overlap for each subject for a total batch size of 256 windows split across two Tesla V100 SXM2 with 16GB of memory each. In this way the model sees a random 24hrs of data per recording in each epoch. The 50-minute context window was chosen for memory constraints with the given batch size. Pretraining was run for 20 epochs and took ~24hrs. A base learning rate of $10^{-3}$ with one-epoch linear learning rate scaling for burn-in followed by cosine annealing with warm restarts every epoch. Unless otherwise specified a masking rate of 60% was used and the reconstruction loss was calculated between the input and the reconstruction of the masked patches.

## E. Fine-tuning

For fine-tuning an 80:20 train-test split was used. The train set was further divided into 5 cross-validation (CV) folds with an 80:20 train-validation split each. During fine-tuning, the full 24-hour recordings were loaded into memory and divided into 50-minute windows with 45 minutes of overlap. This served as a form of data augmentation with the limited scale of fine-tuning data. Fine-tuning was done using a flat learning rate of $2\times10^{-3}$ with a batch size of 512 for 75 epochs on a single Tesla V100 SXM2 with 16GB of memory and took ~1 hour. Only the classification head was optimized during finetuning.

## F. Loss Functions

For self-supervised pretraining we explored three different loss components, including mean squared error (MSE) and two frequency-based loss components. The MSE loss is typically used for image and time series MAEs relying on signal reconstruction [2], [13]. The first frequency-based loss function was the log-scale mean-magnitude (LMM) loss (Fig. 1b) based on a log-scale magnitude loss introduced for audio signals in [20], with the change that we average the spectrogram in the time dimension for each 10-second patch. For the 3-axis accelerometry signals the STFT was applied axis-wise on the inputs and the reconstructions as:

$$U(x) = \log\left(\max(\mathrm{mean}_{time}(|\mathrm{STFT}(x)|), \epsilon)\right), \quad (1)$$

where $\epsilon = 10^{-1}$. For a single input patch $i$ from axis $j$ $x_{ij}$ and corresponding reconstruction $\hat{x}_{ij}$, the LMM loss is computed as:

$$\mathrm{LMM}(x_{ij}, \hat{x}_{ij}) = \mathrm{MSE}\left(U(x_{ij}), U(\hat{x}_{ij})\right) \quad (2)$$

Each patch was zero-padded on both sides so that the $m^{th}$ FFT window was centered on the $m \times H^{th}$ sample where $H$ is the hop size of the FFT window. The Hann-window function was applied to each window before the FFT to mitigate edge effects. We tested the LMM loss in combination with the MSE loss since we discarded any intra-patch temporal information in averaging the magnitude of the STFT windows and not using phase information. The combined loss is defined as:

$$\mathcal{L}_1(x_{ij}, \hat{x}_{ij}) = w_{LMM}\mathrm{LMM}(x_{ij}, \hat{x}_{ij}) + w_{MSE}\mathrm{MSE}(x_{ij}, \hat{x}_{ij}) \quad (3)$$

We also considered a loss on the time-variance of the spectral magnitudes as a way of introducing intra-patch temporal awareness without exact phase information. We denote this as the log-scale magnitude variance (LMV) loss, which was computed as:

$$V(x) = \log(\max(\mathrm{var}_{time}(|\mathrm{STFT}(x)|), \epsilon)), \quad (4)$$

$$\mathrm{LMV}(x_{ij}, \hat{x}_{ij}) = \mathrm{MSE}\left(V(x_{ij}), V(\hat{x}_{ij})\right). \quad (5)$$

We test this in combination with the LMM loss:

$$\mathcal{L}_2(x_{ij}, \hat{x}_{ij}) = w_{LMM}\mathrm{LMM}(x_{ij}, \hat{x}_{ij}) + w_{LMV}\mathrm{LMV}(x_{ij}, \hat{x}_{ij}), \quad (6)$$

The reconstruction losses are only applied to masked input patches. Thus, for all 3 axes $A$ full input signal $X$ and its reconstruction $\hat{X}$ divided into $L$ patches:

$$\mathcal{L}(X, \hat{X}) = \frac{1}{\sum_{i=0}^{L} M_i} \sum_{i=0}^{L-1} M_i \sum_{j=0}^{A-1} \mathcal{L}(x_{ij}, \hat{x}_{ij}), \quad (7)$$

where $M$ is a binary vector of length $L$ with $M_i = 1$ indicating that the $i^{th}$ input patch was masked.

Fine-tuning for classifying human activities was done using the weighted cross entropy loss. Which for a single patch was computed as

$$\mathcal{L}_{CE}(y, \hat{y}) = -\sum_{i=0}^{C-1} w_i y_i \log\left(\frac{\exp(\hat{y}_i)}{\sum_{c=0}^{C-1}\exp(\hat{y}_c)}\right) \mathbf{1}(i \neq g), \quad (8)$$

where $y$, $\hat{y}$, and $w$ are vectors of length $C$, $C$ being the number of classes, $y$ is binary vector with $y_i = 1$ when $i$ is the index of the observed class and zero otherwise, $\hat{y}$ contains the output logits of each class, and $w$ is a vector of weights for each class corresponding to the inverse of class prevalence, $g$ is the class index corresponding to a missing label, and $\mathbf{1}(x)$ is the indicator function.

## G. *Evaluation Metrics*

We report the macro-averaged F1 scores over all classes and Cohen's kappa (κ) to evaluate downstream classification performance. These were chosen as complimentary measures as the macro-averaged F1 weighs performance on each class equally, whereas κ measures overall agreement, preferencing majority classes. We report the metrics in percentage for readability. Unless otherwise stated, uncertainty estimates are ± S.E.M. over 5 CV-folds of the Capture-24 train set or over individuals in the benchmarking experiments.

# III. Experiments and Results

Models were pretrained on the full UK Biobank training partition for 20 epochs and performance was evaluated after 75 epochs of fine-tuning on the Capture-24 dataset for classification of bicycling, walking, mixed, vehicle, sit-stand, and sleep, unless otherwise specified. All reported loss values and metrics are for the validation sets. In 5-fold CV the mean metrics of the last 5 epochs for each fold were used. In benchmarking the best performing epoch was used.

## A. *Loss Function Ablations*

In the first experiments, we tested a model pretrained with the first combined loss (3). We vary the number of samples ($N_{FFT}$) in the STFT windows effectively increasing the frequency resolution at higher values of $N_{FFT}$ while reducing the number of windows in each signal patch since the hop length stays fixed at $\frac{N_{FFT}}{2}$. We varied $N_{FFT}$ between 16, 32, 64, and 128 samples (Table 1) corresponding to ~ 0.5, 1, 2, and 4 seconds respectively. First, the MSE and LMM losses were investigated separately. The best performances for the LMM loss were observed with 32 and 64 samples $N_{FFT}$. The 16-sample window performed worst of the window sizes. However, the LMM loss outperformed the MSE loss at all values of $N_{FFT}$. The combined loss (3) with $w_{MSE} = 0.1$ and $w_{LMM} = 1$ was also tested for possible interactions with the MSE component and $N_{FFT}$. We found that this boosted performance for the 16- and 128-sample windows. We moved forward with a 32-sample STFT window, rather than 64 samples, due to higher computational efficiency of shorter window lengths. Next, we tested the second combined loss (6), varying the weight on the magnitude variance loss, with $w_{LMV} \in \{0,\ 0.01,\ 0.1,\ 1\}$ while keeping $w_{LMM} = 1$ (Table 2). No clear trend was observed for $w_{LMV}$ with the worst performance observed $w_{LMV} = 0.01$ and the best with $w_{LMV} = 0$. Therefore, the LMV loss was not included in further experiments.

TABLE I. FFT Window Sizes

| $w_{LMM}$ | $w_{MSE}$ | $N_{FFT}$ | Macro F1 | Kappa |
|---|---|---|---|---|
| 0 | 1 | - | 58.1 ± 1.2 | 65.1 ± 1.2 |
| 1 | 0 | 16 | 69.9 ± 1.2 | 72.8 ± 0.9 |
| | | 32 | 77.9 ± 1.3 | **80.2 ± 1.0** |
| | | 64 | **78.1 ± 1.2** | 79.7 ± 0.8 |
| | | 128 | 75.3 ± 1.1 | 78.3 ± 0.7 |
| | 0.1 | 16 | 72.7 ± 1.1 | 75.1 ± 0.7 |
| | | 32 | 77.4 ± 1.3 | 79.7 ± 1.0 |
| | | 64 | 76.6 ± 1.1 | 79.2 ± 0.8 |
| | | 128 | 76.0 ± 1.1 | 78.2 ± 0.8 |

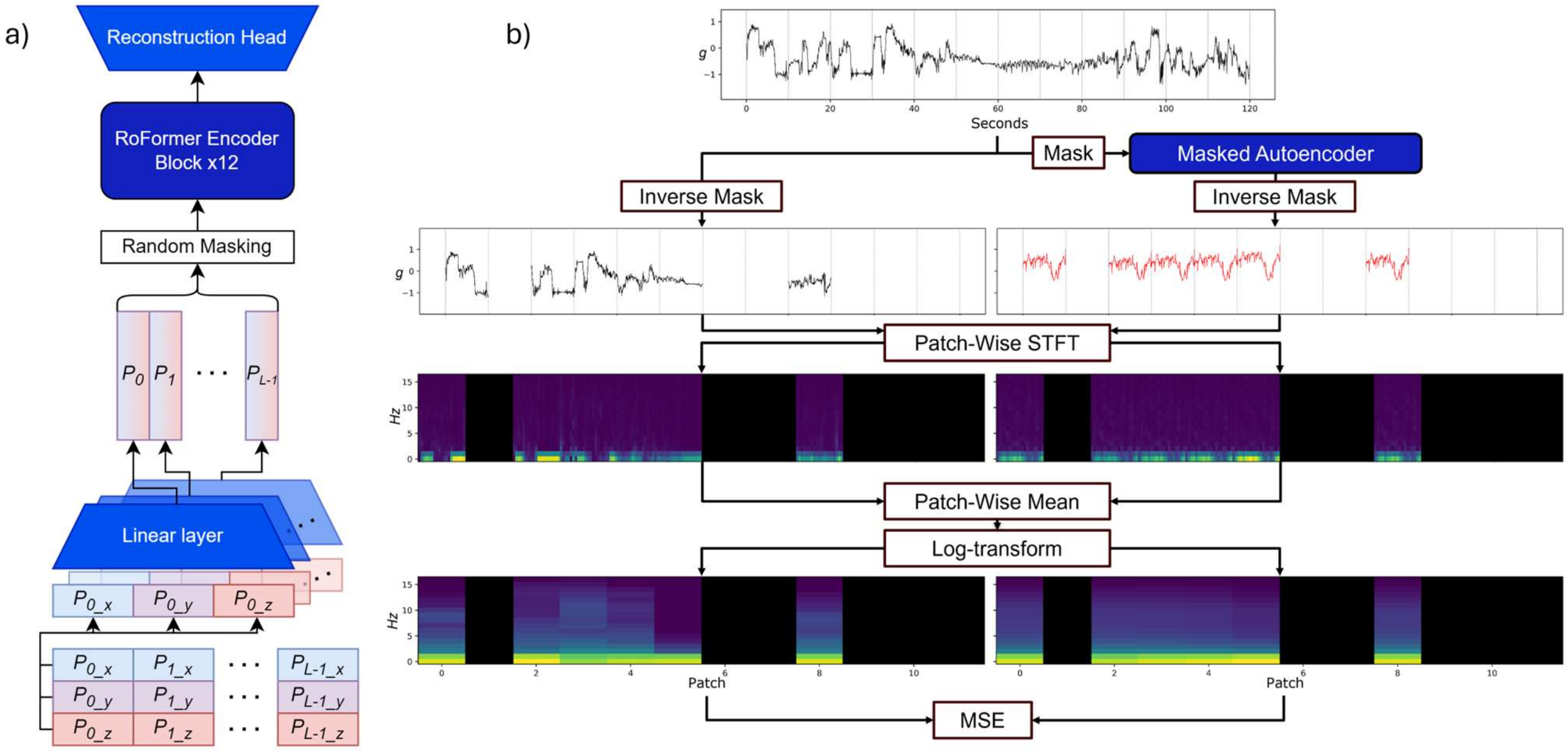

Figure 1. **Model Pretraining Architecture and Log-Scale Mean Magnitude Loss**
**a)** Architecture of the transformer masked autoencoder (MAE) during pretraining. During finetuning the masking block is removed, and the reconstruction head replaced with a classification head. **b)** Illustration of the log-scale mean magnitude (LMM) loss for a single axis of the tri-axial input. First, the patched input (top, horizontal lines denote 10s patches) is passed through the MAE. Then the application of the inverse mask keeps only the patches that were masked in the MAE. Next, magnitude spectrograms for each patch are computed, discarding phases, followed by temporal averaging and log-transformation. Finally, the MSE loss is calculated between these patch-wise log-scale mean spectrograms.

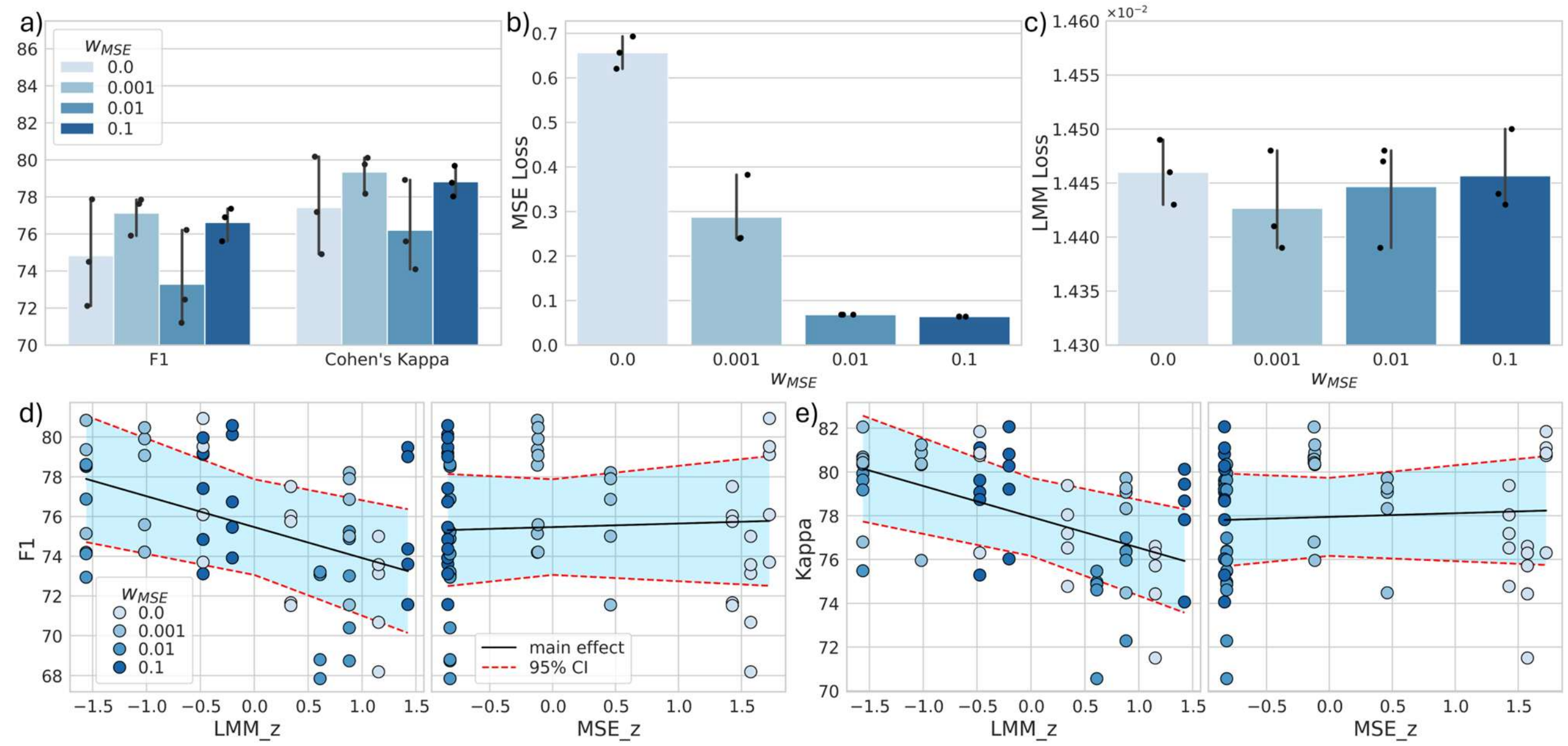

Figure 2. **Pretraining Convergence and Downstream Performance**
**a-c)** Effect of different values of $w_{MSE}$ on **a)** finetuning HAR performance, **b)** pretraining MSE loss, and **c)** pretraining LMM loss. Points are means of 5 CV runs and error bars show ±95% CI. **d, e)** Regression plots showing the association between convergence of the pretrain loss in z-scores (LMM_z and MSE_z) and **d)** downstream F1 or **e)** Cohen's Kappa. Black lines are the main effect of a linear mixed-model regression, with the CV fold as a random effect. For each pretraining a point for each fold is shown.

### *B. Pretraining Convergence*

Next, we assessed the convergence stability of the combined pretraining loss and the effect of pretraining convergence on downstream HAR performance. With $w_{MSE} \in \{0, 0.001, 0.01, 0.1\}$ and $w_{LMM} = 1$, the pretraining was repeated with three different random seeds for each value of $w_{MSE}$. First, we evaluated group differences in downstream performance (Fig. 2a). We did not observe any statistically significant difference in classification performance between any pairs of $w_{MSE}$ when comparing the mean 5-fold CV performance of each pretraining. We then explored the effect of $w_{MSE}$ on the convergence of the MSE loss and the LMM loss components. We report the unweighted values of the MSE loss for comparability, including when $w_{MSE} = 0$. Unsurprisingly, increasing the value of $w_{MSE}$ lead to significantly lower values of the MSE loss (Fig. 2b) with the two highest weights 0.01 and 0.1 leading to very stable convergence. However, no significant effect was observed for $w_{MSE}$ on the value of the LMM loss (Fig. 2c) and a similar spread was observed for all values of $w_{MSE}$. To examine associations between pretraining convergence and downstream classification performance, we fit linear mixed-effects models for F1 (Fig 2d) or κ (Fig 2e) as dependent variables and both z-scored pretraining losses as independent variables. Z-scoring was done for easy comparability across scales. We found that decreasing values of the LMM loss were significantly associated with higher downstream performance ($p < 0.001$) for both F1 and κ, while no significant association was found for the MSE loss. However, some of the best performing models were trained with $w_{MSE}$ of 0.001 or 0.1 (Fig 2a). While this might be due to instability in the pretraining convergence we included the best models pretrained with $w_{LMM} = 1$ and $w_{MSE} = 0$ or 0.001 in the following experiments, referred to as $MAE_{LMM}$ and $MAE_{LMM+MSE}$, respectively, moving forward, for simpler notation.

TABLE II. LMV LOSS

| $w_{LMM}$ | $w_{LMV}$ | Macro F1 | Kappa |
|---|---|---|---|
| 1 | **0** | **77.9 ± 1.3** | **80.2 ± 1.0** |
| | 0.01 | 66.2 ± 1.2 | 68.7 ± 1.4 |
| | 0.1 | 74.4 ± 1.2 | 76.7 ± 0.7 |
| | 1 | 73.0 ± 1.3 | 76.1 ± 0.9 |

### *C. Masking Rate*

In these experiments we did a small grid search for the masking rate used during pretraining with values of 0.3, 0.6, 0.75 and 0.9. This was done with both the $MAE_{LMM}$ and $MAE_{LMM+MSE}$ models. Both with and without the MSE component the best performance was achieved with the initial masking rate of 60% (Table 3). A steady decrease in performance was observed at higher and lower values for $MAE_{LMM}$. For $MAE_{LMM+MSE}$ pretraining instability caused low performance at 75% masking rate with a surprisingly high performance at 90%. It is unclear if the comparatively low performance of $MAE_{LMM}$ at 90% is due to similar convergence instability or if the MSE component makes the model more robust to high masking rates.

TABLE III. MASKING RATES

| Model type | Masking rate | Macro F1 | Kappa |
|---|---|---|---|
| $MAE_{LMM}$ | 0.3 | 75.5 ± 1.4 | 78.1 ± 1.0 |
| | **0.6** | **77.9 ± 1.3** | **80.2 ± 1.0** |
| | 0.75 | 65.7 ± 1.3 | 67.8 ± 1.0 |
| | 0.9 | 53.6 ± 1.2 | 51.0 ± 1.4 |
| $MAE_{LMM+MSE}$ | 0.3 | 73.7 ± 1.2 | 76.4 ± 0.9 |
| | **0.6** | **77.6 ± 1.3** | **80.1 ± 0.9** |
| | 0.75 | 50.0 ± 1.7 | 50.1 ± 1.6 |
| | 0.9 | 73.0 ± 1.5 | 75.7 ± 0.9 |

## D. Representations and Performance

UMAP was used to visualize the representations learned by The $MAE_{LMM}$ and $MAE_{LMM+MSE}$ models, as well as one trained with only the MSE loss ($MAE_{MSE}$) and compared them to their confusion matrices (Fig. 3). The pretrained encoders were used to generate non-overlapping embeddings of the Capture-24 data. In the UMAP projections of $MAE_{LMM}$ (Fig. 3a) we saw good separation between classes except ‘mixed’, walking’, and a subset of ‘sit-stand’. This overlap was partially expected as the ‘mixed’ class includes activities that involve walking. The ‘bicycling’ class was strongly separated from the rest. These observations align well with the confusion matrix (Fig. 3b), where we see most confusion between ‘walking’ and mixed’, and ‘sit-stand’ to a lesser degree, while ‘bicycling’ and ‘sleep’ are almost perfectly classified. The $MAE_{MSE}$ embeddings generally displayed a much weaker grouping of the classes with two distinct clusters appearing in the UMAP plot which both seemed to contain most classes. The confusion matrix revealed increased confusion between ‘sit-stand’ and ‘vehicle’ compared to the LMM-pretrained models but surprisingly strong performance for ‘sleep’ despite no strong clustering of the class in the embedding space. The combined $MAE_{LMM+MSE}$ model showed very similar embedding patterns to $MAE_{LMM}$ with slightly less sharply delineated groupings, which did not impair performance for any specific class.

## E. Data Volume Dependency

We assessed the sensitivity of the models on the amount of unlabeled data seen during pretraining by varying the percentage of the total UK biobank training set used between 20, 40, 60, 80, and 100 percent. For the reduced amounts of training data, the number of pretraining epochs was increased to keep the total number of training steps constant. The reduced training sets are subsets of all the larger sets and the validation set remained the same. This was done with $MAE_{LMM}$ and $MAE_{LMM+MSE}$. For both models we observed that the best performance was achieved with 100 percent of the training data (Fig. 4b). However, we did not see a steady percent of the pretraining data compared to both 60 and 20 percent for both models. We also note that we did not observe a performance plateau suggesting that we might see a performance increase with even more pretraining data. Next, we evaluated the sensitivity of the models pretrained on the full pretraining sample to the amount of fine-tuning data (Fig 4a). The total number of training steps were kept constant by increasing the number of epochs. Both $MAE_{LMM}$ and $MAE_{LMM+MSE}$ showed relatively stable performance when reducing the percentage of training data used from each fold, with a ~1% difference in F1 between 100 and 10%. Surprisingly, the best performance was observed when using 40% of the training data for both models. This might be

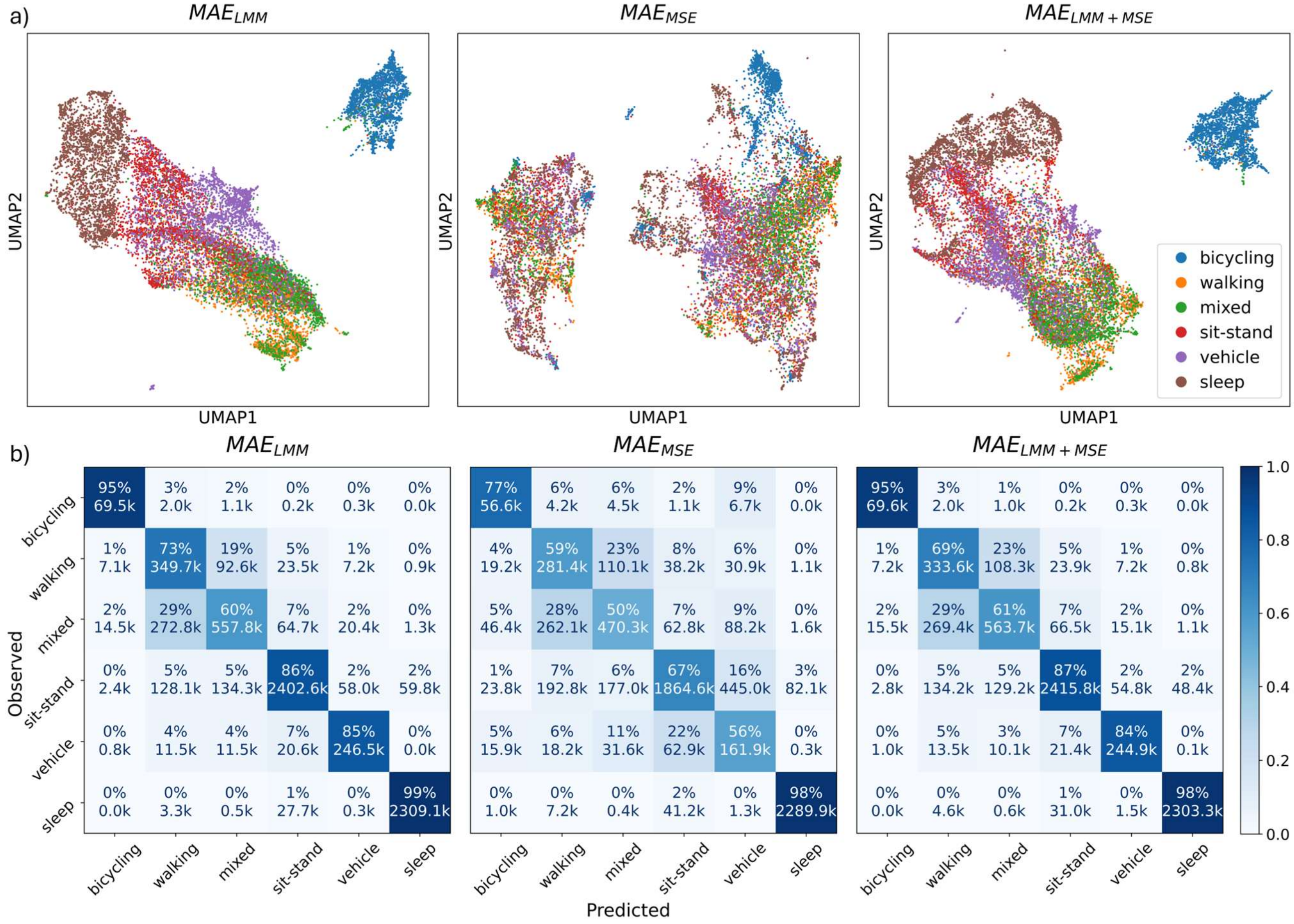

Figure 3. **Behavior Grouping in Latent Spaces and Model Confusion**

**a)** UMAP projections of embeddings from $MAE_{LMM}$, $MAE_{MSE}$, and $MAE_{LMM+MSE}$, respectively. **b)** Confusion matrices for the same models. Top numbers indicate percentage out of total observations (row-wise), the same is true for color intensity. Bottom numbers are absolute numbers in thousands. Predictions are on respective validation sets aggregated from all cross-validation splits.

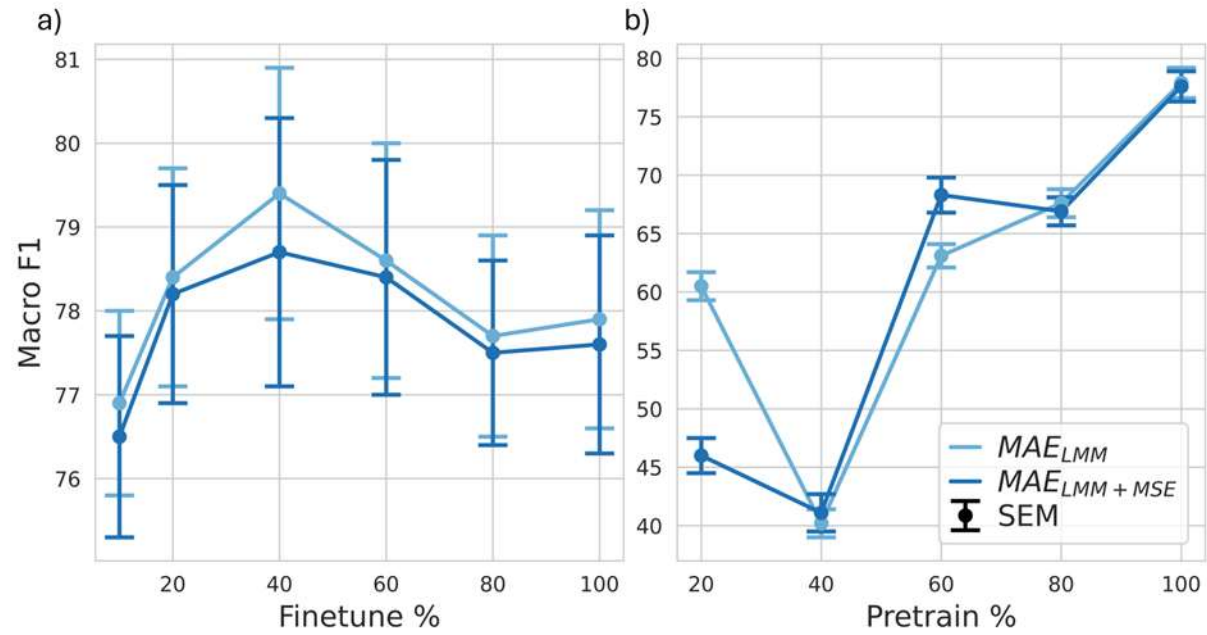

Figure 4. **Data Volume and HAR performance**
**a)** F1 score vs. percent of training data used from each fine-tuning fold. **b)** F1 score vs. percent of pretraining data used.

explained by the weighting of the classes in the CE loss (8) changing with the prevalence of individual labels in the subsets. This could in principle lead to more optimal class weights for the validation sets.

### *F. Benchmarking*

Here, we benchmark our MAE models against HARNet10, the current state-of-the-art [1] in HAR from wrist-worn accelerometer data. The MAE models were fine-tuned with either only a linear layer or 2 short-term memory (LSTM) layers and a linear layer. HARNet10 was fine-tuned with a multi-layer perceptron (MLP) classification head as in [1] or 2 LSTM layers and an MLP. The LSTM fine-tuning was done with 50-minute context windows since the ResNet-based HARNet10 takes in only one 10-second segment at a time while our transformer-based models are sequence-aware with 50-minute context windows. The MAE models fine-tuned with LSTM layers were also included for fair comparison in terms of number of trainable parameters during finetuning. Models were fine-tuned on the entire Capture-24 train set for 75 epochs and we report subject-wise evaluation metrics on the test set from the best epoch (Table 6). HARNet10, with only an MLP head was only trained for 10 epochs as it takes ~10x more steps per epoch, even with a batch size of 4096 compared to the sequence models with batch sizes of 512. When fine-tuning just a linear layer $MAE_{LMM}$ and $MAE_{LMM+MSE}$ had ~10% higher subject-wise F1 scores compared to HARNet10 with just an MLP (Table 6). However, when fine-tuning with LSTM layers HARNet10 obtained a slightly higher F1, with increases of 1.3 and 0.4% compared to $MAE_{LMM}$ and $MAE_{LMM+MSE}$, due to a large 14.9% increase in F1 compared to just the MLP classifier, while F1 increases of only ~4% were observed for the MAE models pretrained with the LMM loss. The $MAE_{MSE}$ model performed worst of the pretrained models with a linear probing performance on par with the transformer trained from scratch and 2.9% lower than HARNet10 and LSTM fine-tuning performance lower than the linear probing performances of the other MAE models. The transformer model trained from scratch was identical in architecture to our MAE models.

TABLE IV. BENCHMARKING

| Model | Fine-tuning Method | Subject-wise Macro F1 | Subject-wise Kappa |
|---|---|---|---|
| $MAE_{LMM}$ | Linear | 73.3 ± 1.2 | 80.0 ± 1.5 |
| | LSTM | 77.1 ± 1.3 | 82.8 ± 1.6 |
| $MAE_{MSE}$ | Linear | 60.6 ± 1.3 | 64.7 ± 1.8 |
| | LSTM | 71.7 ± 1.6 | 78.6 ± 1.8 |
| $MAE_{LMM+MSE}$ | **Linear** | **74.1 ± 0.9** | **80.0 ± 1.5** |
| | LSTM | 78.0 ± 1.3 | 83.9 ± 1.4 |
| $Transformer_{scratch}$ | - | 60.4 ± 1.7 | 66.3 ± 2.1 |
| HARNet10 | MLP | 63.5 ± 1.1 | 65.9 ± 1.7 |
| | **LSTM** | **78.4 ± 1.1** | **84.5 ± 1.4** |

## IV. DISCUSSION

We found that the LMM loss was key to pretraining the MAE model while we did not find that including any intra-patch temporal information through the MSE or LMV losses contributed to learning useful representations for downstream HAR. We found a strong association between the convergence of the LMM loss component and downstream performance while this was not the case for the MSE loss. This suggests stabilizing and improving LMM loss convergence as a promising avenue for future research. It also shows the importance of characterizing the uncertainty in the pretraining convergence when introducing a new self-supervised learning approach, something that is often neglected in current literature, possibly due to constraints on time and GPU capacity. Regarding why the MSE loss proved ineffective for downstream HAR modelling, it seems likely that reconstructing the signal sample by sample in large patches as enforced by the time-domain MSE loss is simply too difficult a task for effective learning. This could also explain the deterioration in performance observed at the highest frequency resolution we tested for the LMM loss. This was our initial motivation for not simply using the full Fourier transformation or the full spectrogram of each patch. On the other hand, a task also needs to be complex enough to facilitate learning of useful features. This could explain reduced performance at the lowest frequency resolutions and masking rates. This explanation fits with the "sweet spots" observed for the frequency resolution and masking rates. The failure of the LMV loss could be due to it being a poor fit for the downstream task, since it accentuates transient events within patches, while the behaviors modeled here are relatively stable at the 10-second scale. In line with the above considerations, the success of the LMM loss for this HAR task is probably a combination of alignment with the downstream task, i.e. the importance of frequency magnitudes, compared to phase, for distinguishing behaviors, and finding a middle-ground of task difficulty. Here, the log-term was used to decrease the relative importance of high-power low-frequency components. Future work could investigate the importances of reconstructing different frequency bands. Additionally, future work could explore self-supervised learning tasks in the self-distillation family such as DINO [28], which focus on the latent representations rather than sample-wise fluctuations, which might afford benefits similar to those of the LMM loss.

In our benchmarking experiments we found that our LMM trained MAE models outperformed HARNet10 when training a simple classifier, but this could be due to the quality of the learned representations or the sequence-aware nature of our model. We included the LSTM classifier to test this and found that it removed the performance gap suggesting similar quality of the individual embeddings. However, it is important to note that the advantage brought by sequence modelling in our MAE models can only be due to detection of natural state transitions in the data, since no sequence-aware training is done with labeled data, while the LSTM is allowed to learn sequential patterns in the labels which could affect generalization to other datasets. Furthermore, training HARNet10 with the LSTM layers increased GPU memory use about 8x and reduced training speed by half compared to the transformer models with LSTM layers. This showcases the advantages of performing sequence modelling during pretraining.

A key limitation of this work is the limited scope of the downstream task. Testing how this pretraining approach generalizes to free-living HAR in other cohorts and devices will be key to demonstrate the strength of the approach and highlights the need for more such datasets to be collected and made available.

## V. Conclusion

In this work, we introduced masked reconstruction of frequency magnitudes and variances with the LMM and LMV losses for self-supervised pretraining for HAR using high-resolution wrist-worn accelerometry. We demonstrated that pretraining with the LMM loss allows a transformer MAE to achieve state-of-the-art performance for HAR, while adding the LMV loss did not improve performance, but also showed that sequence modelling is not enough to make up for bad representations obtained using the MSE loss alone. This is to our knowledge the first demonstration of the viability of pretraining transformer MAEs on large-scale high-resolution accelerometry data for HAR, highlighting the potential of pretraining sequence-based models for free-living HAR.